\documentclass[useAMS,usenatbib]{mn2e}

\usepackage{graphicx}  
\usepackage{amsmath}
\usepackage{soul}
\usepackage{xcolor}
\setulcolor{red}

\def\aj{AJ}
\def\apj{ApJ}
\def\apjl{ApJ}
\def\apjs{ApJS}
\def\aap{A\&A}
\def\mnras{MNRAS}
\def\nat{Nature}
\def\pasp{PASP}%
\def\pasa{PASA}%

\title[Isolated cE with diffuse component]{An isolated, compact early-type galaxy with a diffuse stellar component: merger origin?\thanks{Based on observations collected at the European Organisation for Astronomical Research in the Southern Hemisphere, Chile (programme 087.B-0841(A))}
}
   
\author[S. Paudel et al]{Sanjaya Paudel$^{1,2}$\thanks{E-mail:
sjy@x-astro.net}, Thorsten Lisker$^{3}$,
K. S. A. Hansson$^{3}$,
and Avon P. Huxor$^{3}$\\
$^{1}$Korea Astronomy and Space Science Institute, Daejeon 305-348, Republic of Korea\\
$^{2}$Laboratoire AIM Paris-Saclay, CEA/IRFU/SAp, 91191 Gif-sur-Yvette Cedex, France\\
$^{3}$Astronomisches Rechen-Institut, Zentrum f\"ur Astronomie der Universit\"at Heidelberg, M\"onchhofstra\ss e 12-14, 69120 Heidelberg,
 Germany}
\begin{document}

\date{\today}

\pagerange{\pageref{firstpage}--\pageref{lastpage}} \pubyear{2014}

\maketitle

\label{firstpage}

\begin{abstract}
The relation between the size and luminosity for both bright and faint early-type galaxies has been repeatedly discussed as a crucial proxy for understanding evolutionary scenarios, as galaxies grow and lose their stellar mass in different physical processes. The class of compact early-type galaxies (cEs) are, however, distinct outliers from this relation and mainly found around massive galaxies in the centres of groups and clusters. The recent discovery of a cE in isolation provided a new opportunity to understand their formation scenario in a different environment. Here, we report the discovery of an isolated cE, CGCG 036-042, using imagery from the Sloan Digital Sky Survey (SDSS). The CGCG 036-042 has an $r$-band absolute magnitude (M$_{r}$) of $-$18.21  mag and a half-light radius (R$_{e}$) of 559 pc. Interestingly, it possesses a faint outer stellar component, which extends up to 10 kpc from its centre and has a nearly constant $r$-band surface brightness of $\sim$26 mag/arcsec$^{2}$. ESO-VLT long-slit spectroscopic data reveal that the simple stellar population (SSP) properties are fairly similar to those of previously identified cEs, with an intermediate-age of 7.15$\pm$1.17 Gyr,  a metallicity  of --0.18$\pm$0.07  dex and a supersolar alpha-element abundance of 0.2$\pm$0.05 dex. The SSP metallicity also shows a significant decline outward from the center, by 0.5 dex over one half light radius. We discuss the possible origin of this galaxy and suggest that it might have originated from a merger of even smaller objects -- a well established scenario for the formation of massive early-type galaxies.
\end{abstract}
\begin{keywords}
galaxy -- galaxy interaction -- dwarf galaxy -- tidal evolution -- star-formation
\end{keywords}

\section{Introduction}
Elliptical galaxies follow an overall relation of decreasing size with decreasing luminosity that is continuous over the whole range of luminosities, but is more complex than a simple power law \citep{Graham03,Ferrarese06,Janz08}. The universality of this relation has been a subject of considerable debate: are there different scaling relations for faint and bright galaxies, or for galaxies of low and high surface brightness \citep{Kormendy85,BinggeliCameron1991,Graham08,Kormendy09,Chen10,Misgeld11}? A continuous relation joining bright and faint galaxies would seem to support the notion of a similar origin of these galaxies, regardless of their luminosity \citep{Graham03,Chilingarian09a}.

At high luminosities, elliptical galaxies (Es) follow a steep size-luminosity relation and are commonly believed to be a product of hierarchical galaxy formation through mergers \citep[e.g.][]{Naab07}. In contrast, the faint low-surface-brightness galaxies classified as dwarf ellipticals (dEs), which exhibit a much shallower relation with large scatter \citep{Janz08}, may have acquired much of their present-day structure from environmental influence through tidal forces and ram pressure \citep{Gunn72,Larson80,Moore96,Boselli06}.

Interestingly, the discovery of a number of very compact, low-mass, non-star-forming elliptical galaxies (cEs) adds even more complexity to the interpretation of the observed scaling relations of early-type galaxies \citep{Chilingarian09}.  Despite having a luminosity similar to the dEs, they clearly deviate from the scaling relations, in that they have a much smaller size and higher effective surface brightness at a given luminosity than the dEs \citep{Chilingarian09}. Interestingly, they seem to fall on the extension of the relation defined by bright ellipticals alone \citep{Misgeld11}. They also possess a much larger velocity dispersion than found in a typical dE of similar luminosity \citep[][see Fig. 1 middle panel]{Chilingarian09}.

 Until the advent of large-scale survey and high-resolution imaging from the Hubble Space Telescope, these compact objects were thought to be extremely rare. M32 and VCC1297 are classical examples of cEs in the nearby universe \citep{Faber73}. The population of known cEs, however, is increasing  \citep{Price09,Chilingarian09}. Most of them are found in dense environments, particularly around massive galaxies. Such proximity to a massive host favours the scenario in which they arise through the tidal stripping of larger galaxies, where the inner bulge component becomes a naked compact galaxy after losing its outer disk component during interaction with its massive host \citep{Bekki01,Choi02}.  Indeed, the discovery of cEs with tidal debris near a massive host confirmed this picture \citep{Huxor11}.

In the absence of such environmental factors, it is therefore naively expected that neither dEs nor cEs can be formed in the field. However, the recent discovery of a cE in isolation \citep{Huxor13} has challenged this classical notion of environmental dependent evolution of compact low-mass, early-type galaxies.  In this regard, we present a new field cE, CGCG 036-042, which is closer than any previously identified, at a redshift of  0.007, whereas the \cite{Huxor13} cE has a redshift of 0.019.  
 We present structural and stellar population properties, derived from the Sloan Digital Sky Survey (SDSS) archival images and new optical spectroscopic data that we have obtained from VLT-FORS2\footnote{Very Large Telescope - FOcal Reducer and low dispersion Spectrograph} observation. Finally, we discuss the possible origin of cEs in such environments and their rare existence.

\section{Identification and Data Analysis}

\subsection{Finding isolated early-type galaxies}

As our primary interest is to find low-mass early-type galaxies in an isolated environment using the extensive sky coverage of the SDSS, we carried out a systematic search for such objects  in the local volume (z $<$ 0.01) in the field, deriving the distance to the nearest massive galaxies in terms of sky projected separation and relative velocities. We define isolated galaxies as those not having any nearby massive galaxies (M$_{B}$ $<$ $-$18\footnote{If not mentioned explicitly, the distances are computed assuming H$_{0}$=71 km/s/Mpc throughout the paper.}) within a projected distance of 700 kpc, and a velocity difference of 700 km/s. The morphological selection has been done with visual inspection of the SDSS colour images.

\begin{figure}
\includegraphics[width=8.5cm]{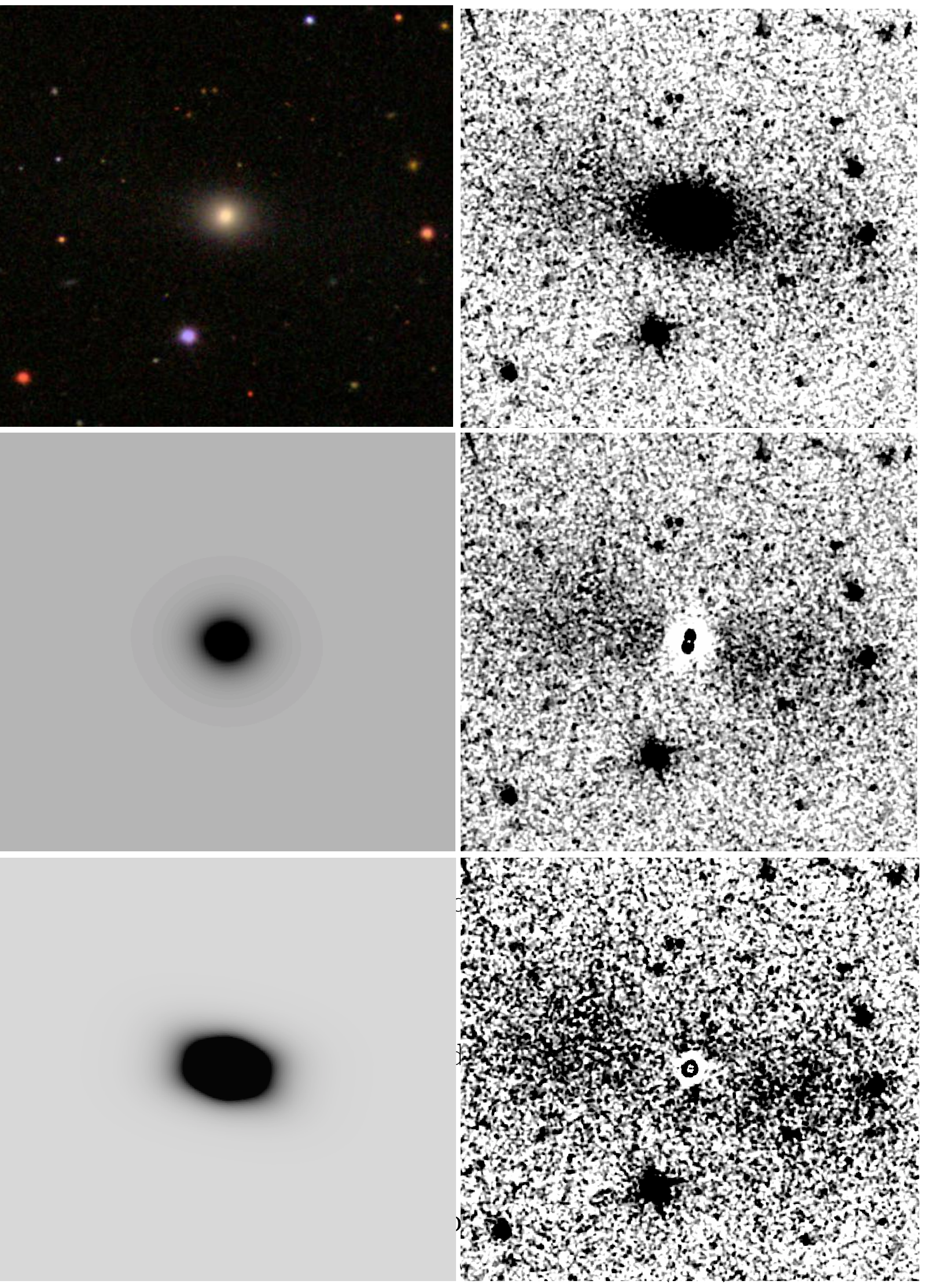}
\caption{SDSS-optical images of CGCG 036-042. Each panel has a size  of 3.4 $\times$ 3.4 arcmin (28 $\times 28$kpc), where north is top and east is left. We show a g-r-i combined true colour image (top left), the r-band image (top right), a one-component GALFIT model (middle left) and its residual (middle right), as well as a two-component GALFIT model (bottom left) and its residual (bottom right).}

\label{gfig}
\end{figure}

At the position RA=10:08:10.32, DEC=+02:27:48.3, and a redshift of z = 0.00687, we found a rare low-mass early-type galaxy, CGCG 036-042,  with an extended outer component (Fig.~\ref{gfig}, top panels). It is located south-west of the NGC 3166 group at an angular distance of 1.67 degrees. Assuming the distance to the NGC 3166 group is 22.6 Mpc\footnote{Based on distance derived  from SN Ia observation, this is consistent with the distance derived from Hubble low for a Virgo + GA infall corrected average redial velocity of group members v$_{r}$ = 1598 km/s  \citep{Waddell12}.}, the  angular separation corresponds to a physical projected distance of 660 kpc. However, the difference in radial velocity between the NGC 3166  group and CGCG 036-042 is 755 km/s, only slightly larger than the relative velocity criterium that we applied to exclude it as a group member, which is  $\sim$10 times greater than the standard deviation of the velocities of known group members \citep{Makarov11}. 

In the following, we take, with no evidence to suggest that CGCG 036-042 is located at the distance of NGC 3166 group, the distance to CGCG 036-042 to be 33.1 Mpc, based on the Hubble law and the  Virgo + the Great Attractor infall corrected radial velocity \citep{Mould00}, and it is thus \emph{not}  a member of NGC 3166 group. In this case, it has an r-band apparent (absolute) brightness of $14.41$ ($-18.21$) mag.

It is notable that we do find a low-mass star-forming galaxy near to CGCG 036-042, at a sky-projected distance of 67.7 kpc with a relative radial velocity less than 100 km/s (RA=10:08:06.94, DEC=+02:20:42.4 \& redshift=0.00655). Adopting the cE's distance, the companion has an absolute magnitude  (M$_{r}$) of $-16.71$ and is thus 1.5 mag fainter than CGCG 036-042.

\subsection{Imaging data analysis}

To perform a detailed image analysis we retrieved archival images from the SDSS-III database \citep{Ahn12}, using the $r$-band image, since it provides a higher signal to noise ratio (S/N) than other bands. CGCG 036-042 is covered by two adjacent pointings of SDSS, but is located right at the edge of both. We therefore needed to combine both fields to enclose the entire galaxy and extended stellar halo into a single image file. Before combination, the sky-backgrounds were subtracted individually in each field, sampling the sky background at 10 independent positions around the galaxy where each individual sky region has an area of 10 $\times$ 10 pixel box. The sky-backgrounds were then computed as an average of sampled sky in each field.

After combining the two pointings, the level of sky-background was cross-checked by comparing the mean sky-background of different subfields\footnote{We first divided the image into two halves (subfields) where each half comes from different pointing and subsequently the sky values were computed for each subfield by sampling the sky background from several independent position, similar to previous sky background sampling in individual SDSS fields.}. We found that the difference in mean sky-background between the subfields is well within the sky-standard deviation of each subfields. The seeing in these fields is rather good, with an $r$-band Point Spread Function (PSF) Full Width Half Maximum (FWHM) of 0.9 arcsec.

\begin{figure}
\includegraphics[width=7.8cm]{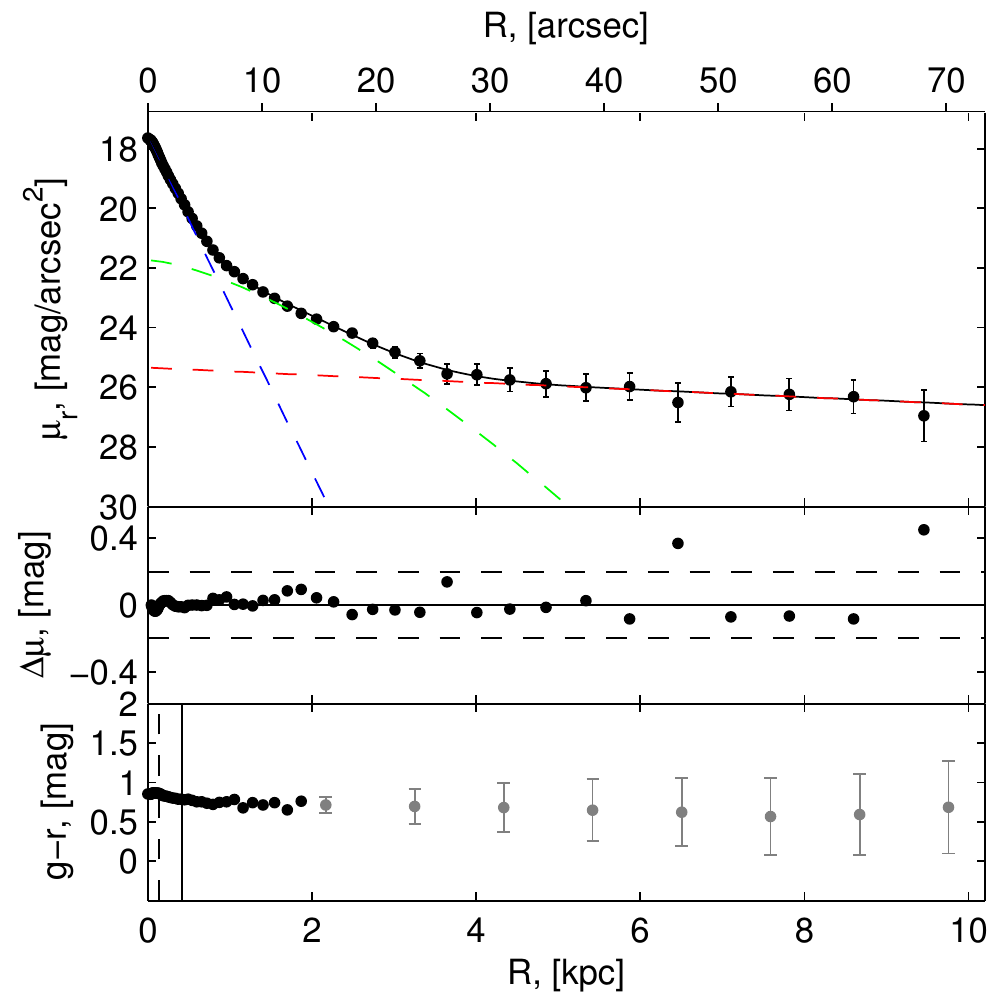}
\caption{ Major-axis r-band light profile fit with a multiple-component S\'ersic model. The outer extended stellar component is modelled with an exponential profile (red dash line) and subtracted from the galaxy main body. The galaxy main body is modeled with a two-component S\'ersic profile, its inner component (blue dashed line) having a significantly smaller effective radius than its outer (green dashed line) component. The black solid line represents the overall model of the galaxy. The residuals -- difference between the overall best fit model (black line) and the data points (black dot) -- about the fit are shown in middle panel. The PSF corrected $g-r$ colour gradient is shown in lower panel. The positions of the dashed and solid vertical lines represent the mean PSF FWHM in the SDSS r-band image and the effective radius of the galaxy, respectively. The colour profile beyond 2kpc (grey) is derived in bins of 20$\times$20 pixels placed outward along the major axis.}
\label{mfig}
\end{figure}

 We used the IRAF task $ellipse$ to extract the galaxy's major-axis light profile, shown in Fig. \ref{mfig}. During the ellipse fitting the centre and position angle were held fixed and the ellipticity was allowed to vary. The centre of the galaxy was calculated using the IRAF task $imcntr$ and the position angle is determined using several iterative runs of $ellipse$ before the final run.

\begin{figure*}
\includegraphics[width=15cm]{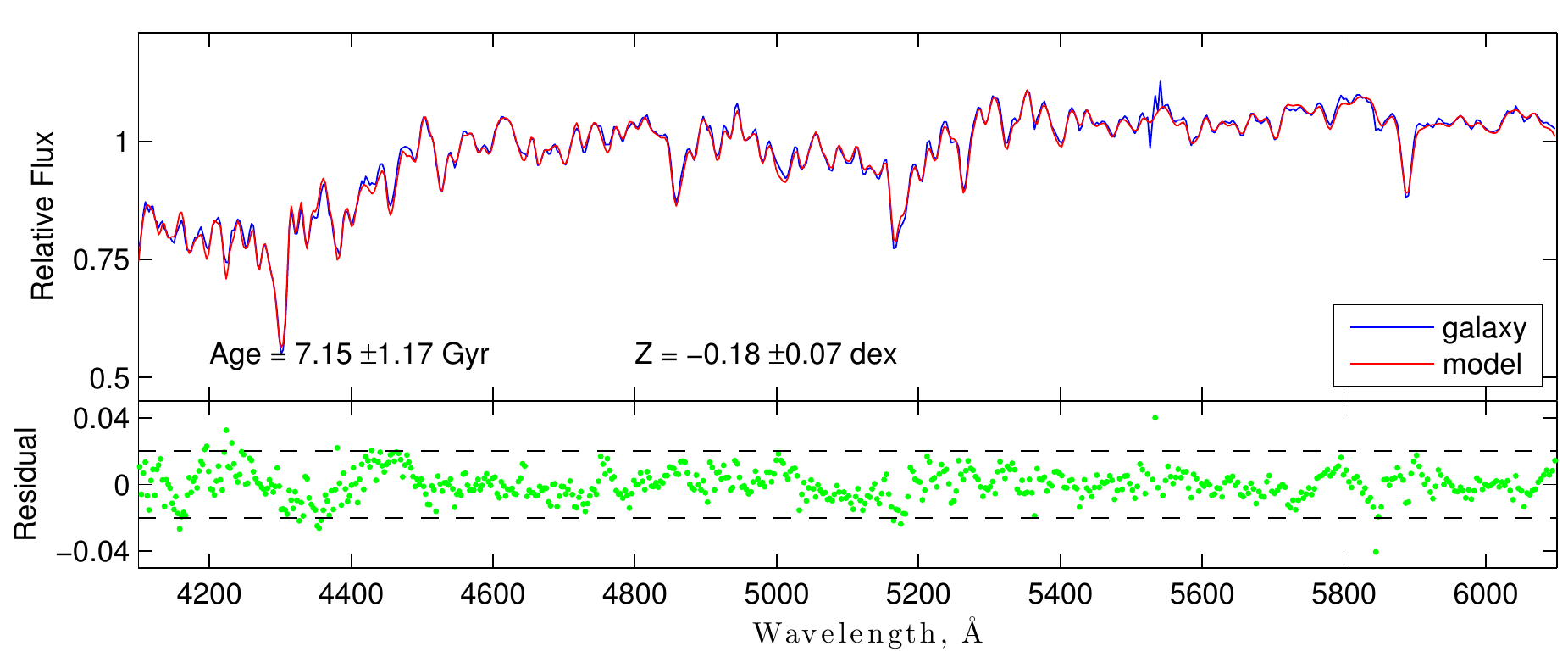}
\caption{The observed spectrum (blue) derived from the light within an aperture of radius 2R$_{e}$ = 7"  along the major axis, together with its best-fit SSP model spectrum (red). The fit residuals are shown in the lower panel. The fit is generally consistent within 2\% of the observed flux.}
\label{fitg}
\end{figure*}

Due to the rather complicated light profile (see Fig. \ref{mfig}, top panel) it is difficult to approximate the observed light distribution with a single parametric profile and then derive the global photometric parameters. The photometric parameters -- half-light radius and total luminosity -- are therefore derived using a non-parametric approach. We follow a similar procedure as described in \cite{Janz08}. The total flux is computed within an aperture of twice the Petrosian radius \citep{Blanton01,Petrosian76} and corrected for the missing light outside of aperture as suggested by \cite{Graham05}, although this correction is only 0.1 mag.

\begin{table}
\begin{tabular}{ccccccc}
\hline
Method & Re$_{in}$ & Re$_{out}$ & $\mu_{in}$ & $\mu_{out}$ & n$_{in}$ & n$_{out}$\\
\hline
IRAF/ellipse & 2.26 & 9.45 & 19.31 & 22.93 & 1.06 & 0.78\\
GALFIT & 2.48 & 11.70 & 18.84 & 23.07 & 1.7 & 1.2\\
\hline
\end{tabular}
\caption{Best-fit model parameters that are obtained from different methods, i.e.\ the fit of the one-dimensional light profile (see Fig.~\ref{mfig}) and the two-dimensional modelling of the light distribution with GALFIT.}
\label{modpartbl}
\end{table}

To model the light distribution with multiple components, two alternative methodologies were used. Firstly, with GALFIT \citep{Peng02}, we performed  two-dimensional modelling of the galaxy image. We derive our PSF model, required by GALFIT, from nearby high-S/N unsaturated stars in the same field. Secondly, we obtain the best fit model parameters from the one-dimensional light profile extracted with IRAF $ellipse$, where a $\chi^{2}-$minimization scheme has been applied. Thereby, we first modelled the extended outer component with an exponential profile, and then fitted a two-component S\'ersic profile to the galaxy (Fig.~\ref{mfig}).  To check that the observed multiple component is not an artifact resulting from sky-background fluctuations, we remodelled the galaxy light profile with an offset of $\pm\sigma$$_{sky}$. For any value of offset, within  $\sigma$$_{sky}$, 
no best-fit single component S\'ersic model is found; and in case of multi-component modelling, the variations in the output model parameters are minimal, for example the variation in the S\'ersic index n is only 10\%.

The colour gradient was derived from the azimuthally-averaged light profiles in the $g$ and $r$-band. However, a different PSF in those bands can produce an artificial colour gradient at the centre of galaxy. Since the $g$-band imaging has worse seeing than the $r$-band (FWHM of 1.23" and 0.92" respectively) --  we first calculated the average FWHM of bright and non-saturated stars around the galaxy in both bands, then degraded the $r$-band image by smoothing it with a Gaussian of $\sigma=0.92$\,pixels,\footnote{The  smoothing applied is slightly different from the expected mathematical value resulting from the squared difference of the PSFs, i.e 0.88 pixels. We found a best match with a smoothing of 0.93 pixels.} and then obtained the light profiles in both bands.

 \subsection{Spectroscopic data analysis}

To perform a detailed analysis of stellar population properties we obtained long-slit spectra along the major-axis using VLT-FORS2\footnote{The data was obtained as a part of ESO VLT proposal ,``Cluster, group and isolated early-type dwarfs: Linking formation paths to stellar population characteristics", 087.B-0841(A).}. We kept a similar instrumental set up as in  \cite{Paudel10}, where we specifically aimed to perform a study of stellar populations from low-resolution spectroscopy. We used the V300 Grism that provides a final instrumental resolution of 11 \AA{} for a slit width of 1". We therefore do not attempt to derive any kinematic information from it. The total exposure time was 1500 sec. Observations were carried out in service mode. Standard stars were observed in the same night as the galaxy.

To reduce and calibrate the VLT-FORS2 long-slit spectroscopic data we followed the procedure of \cite{Paudel10}. The one-dimensional wavelength spectra were extracted by summing the flux in the two-dimensional frame along the spatial direction, using an aperture of suitable width according to our interest of analysis, see below.

We then derived SSP parameters by fitting the extracted one dimensional full spectrum of wavelength range 4200 to 6000 \AA{} to simple stellar population models from \cite{Vazdekis10}. For this purpose, we used the publicly available full spectrum fitting tool ULySS\footnote{http://ulyss.univ-lyon1.fr} \citep{Koleva09}. This method allow us to use maximum information available in the observed spectrum of galaxy during the comparison to the model prediction. The quality of model comparison with observed spectrum of galaxy is shown in Figure \ref{fitg}, where the observed spectrum perfectly matches within 2\%  of the flux. We do not observe any major deviation in any specific line. A minute deviation around the Mg absorption line (at 5177 \AA), is observed, indicating an inconsistency with the applied model which has a fixed value of alpha-element abundance [$\alpha$/Fe]. This is confirmed by our independent measurement by the method of Lick indices,  where the use of an SSP model that has variable [$\alpha$/Fe] yields a super solar value of  [$\alpha$/Fe] (see below).

\section{Results}
\label{sec:results}

Although compact as seen in the colour image, the SDSS r-band image also reveals a faint outer stellar component which extends up to 10 kpc far from the centre with an average surface brightness of 26 mag/arcsec$^{2}$. This can be prominently seen in Fig. \ref{gfig}, where we have shown SDSS colour image (left) and SDSS r-band grey-scale image (right) in the top panel.

The results of the two-dimensional modelling using GALFIT are shown in the lower two panels of Figure \ref{gfig}; its resulting parameters are given in Table~\ref{modpartbl}.  We have modeled the galaxy -- excluding the faint outer component -- with only a single S\'ersic component, and also independently with two S\'ersic components. Visual inspection of the residual images, obtained through subtracting the best fit model from the original galaxy image, clearly reveals the complexity in the galaxy light distribution. It can easily be recognized that the galaxy light is better represented by the two component model, with S\'ersic indices of 1.7 and 1.2 for inner and outer component, respectively. In addition to these components, the diffuse outer component with its low surface brightness can be distinctively identified in the residual images for both models.

The derived one-dimensional surface brightness profile along the major axis of the galaxy from the SDSS-$r$ band image is shown in Figure \ref{mfig}, where it is apparent that the observed light profile can not be approximated with a simple S\'ersic function. As noted above, we model the light profile with an exponential profile for the extended stellar component (red dashed line) and a two-component S\'ersic profile for the main body of the galaxy (blue and green dashed lines). The inner core is more compact than the exterior component, having an effective radius of 2.26".  This is slightly smaller than the overall half-light radius of this galaxy. The best fit parameters are $\mu_{e,inner}$ = 19.31 mag/arcsec$^{2}$,  R$_{e,inner}$ = 2.26",  $\mu_{e,outer}$ = 22.93 mag/arcsec$^{2}$ and R$_{e,outer}$ = 9.45".

The resulting fit parameters, given in Table \ref{modpartbl}, differ for the two methods. In particular, both the inner and outer profile are found to be steeper with GALFIT. However, this was to be expected, as we did not separately model the extended outer component with GALFIT. Therefore, the light contribution of the extended component directly influences the GALFIT model (cf.\ Fig.~\ref{gfig}, bottom left). We used this to confirm the existence of the extended outer component as a \emph{separate} component, which can still be clearly seen in the residuals after subtracting the GALFIT model (Fig.~\ref{gfig}, bottom right). After inward extrapolation of the one-dimensional profile, the light fraction contained in this component is $\sim$ 20\% of the total galaxy light.

 \begin{table*}
 \caption{ Global properties of isolated cEs from this study (P14) and \citet[H13]{Huxor13}. Using the value of H$_{0}$ = 71 km/s-Mpc, for the Virgo + the Great attractor infall corrected radial velocity v$_{r}$ = 2353 km/s, we obtain the distance to our candidate cE and scale factor 33.14 Mpc and 0.158 kpc/", respectively. The galactic extinction is corrected from \citet{Schlafly11}. The values of age, metallicity and alpha-element abundant [ratio] are derived from comparing the observed spectrum with the prediction from simple stellar populations, see text. }
\begin{tabular}{lccccccccccccc}
\hline
Galaxy                  &  M$_{r}$ &  $g-r$  & R$_{e}$ & z & Age & [Z/H]  & [$\alpha$/Fe] & $\sigma$ & $<\mu_{e,r}$$>$\\
    &   mag &  mag  & kpc&   & Gyr &    dex& dex & km/s & mag/arcsec$^{2}$\\
\hline
CGCG 036-042 (P14)    &   $-$18.21 & 0.77 &    0.559 &   0.007&7.15$\pm$1.17&  $-$0.18$\pm$0.07 & 0.20$\pm$0.05 & 90$\pm$8 & 19.64 \\   
\hline
                                  &  &  &  & & & [Fe/H]  & [Mg/Fe] & & \\
J094729.24+141245.3 (H13)         &   $-$18.78   &  0.74 &    0.499 &   0.019 &9.21$\pm$3.00    &  $-$0.19$\pm$0.16 & 0.09$\pm$0.14 &105$\pm$6 & 19.05\\
\hline
\end{tabular}
\label{ptb}
\end{table*}

 The inferred values of photometric parameters are listed in Table \ref{ptb}. Although the CGCG 036-042 exhibits several structural features -- extended outer component, inner and outer core -- overall it has a relatively small half-light radius,  559 pc, half that of typical dEs in the Virgo cluster \citep{Janz08}. The mean surface brightness within the half-light radius, $<$$\mu_{e,r}$$>$, is 19.64 mag/arcsec$^{2}$ and the dispersion velocity --from SDSS spectroscopy-- is fairly large for its luminosity, i.e $\sim$ 90 km/s, consistent with the typical properties of cEs \citep{Chilingarian09}.

 \begin{figure}
\includegraphics[width=8.5cm]{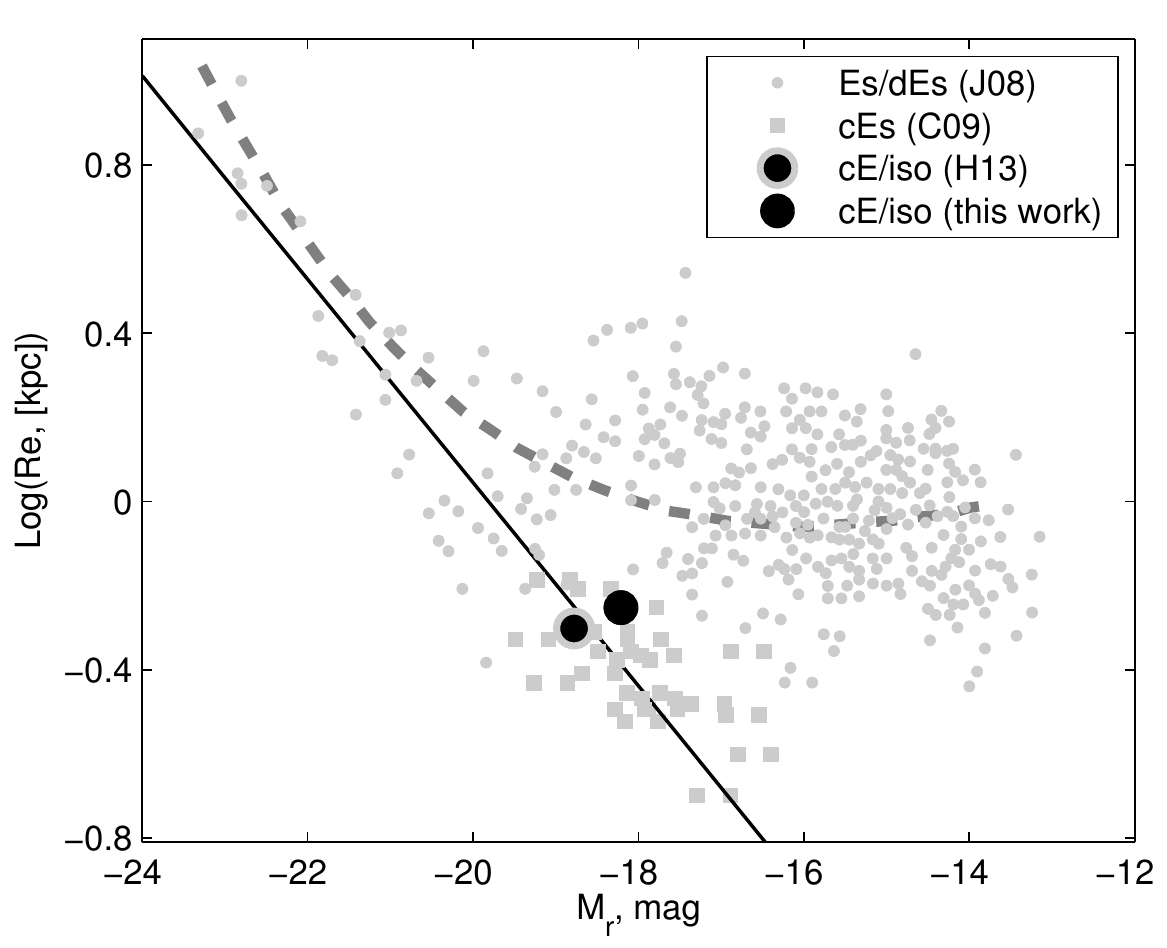}
\caption{Size-luminosity relation for early-type galaxies. Data for Es/dEs (grey dots) are from \citet{Janz08}, while data for cEs (grey squares) are from \citet{Chilingarian09}. A prediction for the scaling relation of Es/dEs according to \citep{Graham03,Janz08} is shown as curved dashed line. The isolated cEs are shown with black filled circles, where the one with a grey surrounding circle represents the cE of \citet{Huxor13}. The solid black line represents the linear best fit to bright early-type galaxies with M$_{r}  < -19.2$ mag. Note that the $r-$band magnitude of the \citet{Chilingarian09} sample was calculated from the B-band absolute magnitude provided in the literature, applying a typical $g-r$ colour of 0.75 mag for the transformation according to \citet{Lupton05}.}
\label{szm}
\end{figure}
 
In Figure \ref{szm}, we show the distribution of sizes and luminosities for early-type galaxies, using the measurements from \citet{Janz08} for  dEs, Es and S0, and \citet{Chilingarian09} for cEs. As indicated in the figure, Es and cEs may follow a common relation between size and luminosity described by a power law (solid line, from a fit to galaxies with $M_{r} < -19.2$ mag). However, based on the size-luminosity relation (dashed line) that results from the observed correlation of central surface brightness and (global) S\'ersic index with luminosity, Es and dEs may also constitute a common population \citep{Graham03}, albeit showing significant systematic offsets at intermediate luminosities (\citealt{Janz08}, also seen in Fig.~\ref{szm}). We can clearly see that our galaxy (filled black circle) falls significantly below the curved size-luminosity relation of Es/dEs, and instead lies in the same region of the plot as the cE sample of \citet{Chilingarian09}, along with the isolated cE of \citet{Huxor13}.

The $g-r$ colour gradient along the major axis of CGCG 036-042 is shown in the bottom panel of Figure \ref{mfig}. We found a weak negative colour gradient in the inner part of the galaxy. The $g-r$ colour index is 0.88 mag at the centre and 0.80 mag at R$_{e}$. A slight negative gradient may continue beyond 2 R$_{e}$, but cannot be confirmed due to the increasing errors.

 \begin{figure}
\includegraphics[width=8cm]{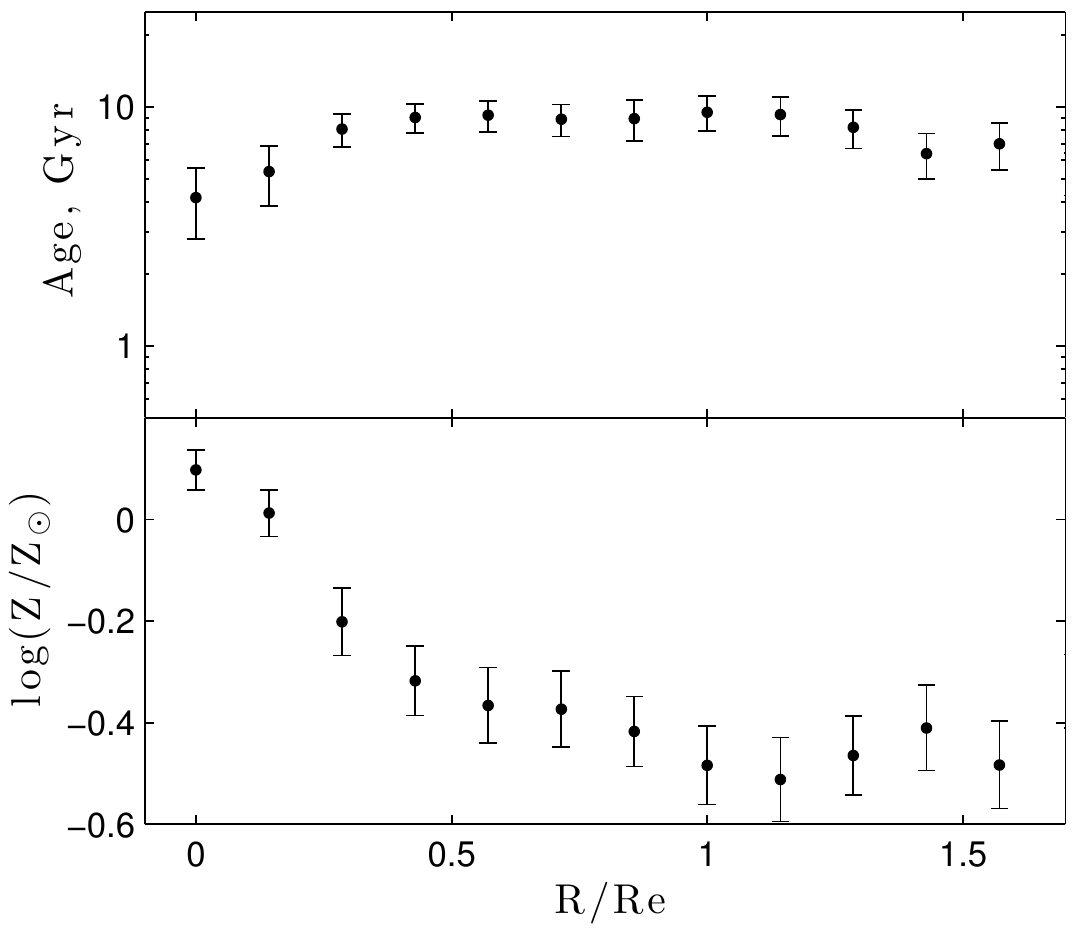}
\caption{Gradient in age and metallicity along the major axis.}
\label{grad}
\end{figure}

The overall stellar population properties of galaxies, i.e. age, metallicity (Z) and alpha abundance ratio, were computed from the one dimensional spectrum extracted using an aperture of $\pm2$\,R$_{e}$ that represents a large part of the galaxy. We found a global age and metallicity of $7.15\pm1.17$ Gyr and $-0.18\pm0.07$ dex respectively (Figure \ref{fitg}). We also derived the spatially resolved profile of age and metallicity along the major axis (see Figure \ref{grad}). While the SSP age increases with distance  from the centre of the galaxy and remains nearly constant at 10 Gyr beyond 0.5 R$_{e}$, it is evident that the metallicity decreases,  from a super-solar value at the centre to $-$0.5 dex at 1 R$_{e}$.

Since there is no SSP model that provides model spectra for a variable [$\alpha$/Fe], we use Lick indices, as in \cite{Paudel10}, to derive the [$\alpha$/Fe] ratio. We measured nine (H$_{\delta}$F,H$\gamma$\,F, Fe4383, H$\beta$, Fe5015, Mgb5177, Fe5270, Fe5335 and Fe5406) well-defined absorption line indices in the Lick system \citep{Burstein84,Trager98} from the extracted one-dimensional spectrum and fitted the models of \cite{Thomas03} to them. We found a super-solar value of [$\alpha$/Fe]$=0.2\pm0.05$ dex, with age and metallicity of 8.9 Gyr and $-$0.24 dex, in very good agreement with the age and metallicity values from the spectral fitting with UlySS.

\begin{table}
\begin{tabular}{lccr}
\hline
Index   & P14  & P14(SDSS)      &   H13           \\
\hline
H$_{\delta}$F &  0.430$\pm$0.136  &  0.613$\pm$0.209  &  0.694$\pm$0.355\\
H$\gamma$F    & -1.055$\pm$0.112  & -1.356$\pm$0.199  & -0.732$\pm$0.339\\ 
Fe4383        &  4.888$\pm$0.224  &  4.527$\pm$0.236  &  4.404$\pm$0.408\\ 
H$\beta$      &  1.699$\pm$0.101  &  2.022$\pm$0.172  &  2.213$\pm$0.282\\ 
Fe5015        &  4.582$\pm$0.211  &  4.710$\pm$0.278  &  4.737$\pm$0.457\\ 
Mgb5177       &  3.669$\pm$0.104  &  3.774$\pm$0.175  &  3.355$\pm$0.292\\ 
Fe5270        &  2.506$\pm$0.114  &  2.738$\pm$0.186  &  3.125$\pm$0.299\\ 
Fe5335        &  1.967$\pm$0.127  &  2.623$\pm$0.190  &  2.713$\pm$0.304\\ 
Fe5406        &  1.508$\pm$0.097  &  1.717$\pm$0.162  &  1.738$\pm$0.261\\
\hline
\end{tabular}
\caption{Measured absorption line strengths in the Lick/IDS system for nine major indices that are used to compare with SSP models. The data for CGCG 036-042 (first column) is obtained from a VLT-FORS2 spectrum. The values in second and third columns were measured and provided by SDSS pipeline \citep{York00} for our and \citet{Huxor13} galaxies, respectively. Note, we used wider aperture of $\pm$7" to extract one-dimensional spectrum from FORS2 data, while the SDSS provides a fiber spectrum of diameter 3" from the centre of galaxies.}
\end{table}

\section{Discussion and Conclusions}
We have presented a rare isolated cE, CGCG 036-042, of which only one other has been identified to date. CGCG 036-042 exhibits similar structural, kinematic and stellar population properties to that of \citet{Huxor13}. However, it remarkably shows a faint stellar component that extends up to 10 kpc from the centre. Considering these evidences, we discuss below the possible formation scenarios of cEs in isolation.

Although dEs and cEs share a luminosity range, cEs exhibit very different structural,  kinematic and stellar population properties \citep{Chilingarian09}. They are not only an outlier in the size-luminosity relation, but also possess relatively high velocity  dispersion and alpha-element abundance ratios compare to normal dEs. A typical dE has a velocity dispersion of less than 50 km/s and near solar  [$\alpha$/Fe], whereas cEs generally exhibit central velocity dispersions of  nearly 100 km/s and a significantly super-solar  [$\alpha$/Fe] \citep{Chilingarian09,Paudel10}.

The branching of the size-luminosity relation was extensively discussed by \cite{Misgeld11}: cEs and the so-called ultra-compact dwarf galaxies (UCDs) seem to rather follow the relation of Es. \citeauthor{Misgeld11} proposed a maximum compactness, i.e.\ maximum density, for any old stellar system.  However, note that the scatter of data points in the dE region blurs the distinction between dEs and cEs \cite[see also][Figure 9]{Janz14}, although the cEs remain consistently at the lower part of the scattered data points as we can see in Figure 4.

High-redshift observations \citep[for example][]{Dokkum08} have revealed that extremely compact early-type galaxies -- but more massive than typical cEs -- are already in place at $z>2$. The formation of these super dense objects has been studied from observational and theoretical perspectives \citep{Gobat13,Hopkins09a}. The general consensus is that dissipational collapse of matter during mergers of gas-rich galaxies leads to the formation of compact Es \citep{Bournaud11}.  
This scenario is also a good candidate for the origin of isolated cEs.  The observed super-solar value of the alpha-element abundance in our cE indicates that the majority of stars were formed in a short burst of star-formation \citep{Thomas05}, consistent with such a scenario. The observed gradient in age and metallicity can also be interpreted as the result of a central starburst after a merger, due to tidal inflow of star-forming gas driven by the interaction \citep{Hopkins09b},  leading to the formation of a dense stellar core \citep{Mihos94}. Furthermore, the presence of fine structure -- stellar bridges, shells and streams -- in early-type galaxies has also been interpreted as the signature of tidal interactions or mergers \citep{Duc11}. The observed extended outer component in our cE may be the consequence of a tidal interaction. Alternatively, it might stem from the accretion of an even lower-mass galaxy \citep[see][]{Rich12}. However, such an accretion origin of the extended component would obviously not help to understand the existence of the cE itself in the first place.

As CGCG 036-042 is located far beyond the nearest massive galaxy and has a large relative velocity, we think it is unlikely  that it might have been disrupted by it.  Such mechanisms are observed very near to the giant host or the centre of a cluster or group of galaxies. For example, both cE systems of \cite{Huxor11} have sky projected separations of less than 20 kpc from a massive galaxy and relative velocities of less than 200 km/s. Moreover, in our extensive search of disrupted dwarf galaxies (\citealt{Paudel13} and Paudel et al.\ in prep.) around massive galaxies in the SDSS data, we find no low-mass galaxies that are visibly in the process of disruption located more than 500 kpc from the massive host or with relative velocities larger than 300 km/s -- in contrast to the cE presented here, with 660 kpc and 755 km/s, respectively.

While we cannot entirely rule out a high speed fly-by of a bulge-dominated disk galaxy, it remains to be shown that this would be effective in destroying the entire outer disk during the very short interaction time, since such interactions have so far mainly be considered for inducing structure in the disk (e.g.\ bar formation, \citealt{Berentzen04}).  In numerical simulation, it has been shown that the probability of a merger origin for low mass galaxies is very small compared to their more massive counterparts. \cite{Lucia06} predict that only one third of their least massive early-type galaxies 
have two progenitors. However, note that this result is primarily for cluster environments and, in contrast, \cite{Klimentowski10} predict that merger of sub-halos is significant  outside of main host halo, and that small satellite galaxies located relatively far from the central galaxy might have a merger origin. Also, on the observational side, the number of known dwarf-dwarf merger candidates have  dramatically increased recently \citep{Amorisco14,Rich12,Penny12}.

The characterisation of environment is not straightforward and the criteria applied to define the various environments: isolated, field, group and cluster are diverse in literature \citep{Karachentsev72,Kuntschner02,Marquez04,Prada03}. Although CGCG 036-042 is located near to the NGC 3166 group in sky projection, it falls  outside of the group's zero-velocity surface  \citep{Makarov11}.
We had applied a simple definition of isolation to select isolated galaxies, -- though our constraints seem more strict than the criteria used to prepare the Catalog of Isolated Galaxies \citep[CIG][]{Karachentsev72}. They use a simple criteria that the galaxy (with a diamter d) should have no companion of size d/4 to 4d within the sky projected distance of 20d. As describe above, we do find a low-mass star-forming galaxy close to our cE, which defies the definition of isolation,  ours and others e.g \cite{Prada03}, but interestingly fulfils the CIG criteria. As  CGCG 036-042 is  outside of NGC 3166 group,  perhaps one can consider it an isolated pair outside the group.

 Moreover, the presence of small companion, if it is truly bounded to CGCG 036-042, might reinforce the our idea of independent evolution of CGCG 036-042 from the nearby group: the alleged pair have to remain bound  during the fly-by of the group, and it is unlikely that the smaller disk galaxy would not suffer any disturbance during this interaction.

Nevertheless, having only slightly more than a quarter of the CGCG 036-042's luminosity and -- based on the companion's bluer colour -- probably less than a fifth of its stellar mass, it is thus unlikely that a strong interaction between these two low-mass galaxies is ongoing. Nonetheless, while the smaller star-forming dwarf shows no signs of disruption in the SDSS data, deeper images are necessary to reveal possible weak interaction signatures.

At the distance beyond the Virgo cluster, it is very hard to detect such compact objects in the seeing-limited imaging data from ground-based facilities. Space-based telescopes with high resolution imaging capability are needed to measure the sizes of these compact galaxies unambiguously, which have effective radii comparable to the seeing of ground base observations. Unfortunately, such observations are expensive and the available data on cEs  was mostly collected serendipitously, with the primary observing targets typically being nearby massive galaxies. This obviously leads to a biased sample, in which cEs are located close to massive galaxies. Thus it is not surprising that the stripping scenario for cE formation has often considered as adequate. \footnote{Although recent work has suggested that the classic cE, M32 -- thought to have been stripped by its interaction with M31, may instead have been intrinsically compact \citep{Dierickx14}.}

Nevertheless, cEs are indeed rare objects in the local universe, including group and cluster environments: only a handful of cEs have been reported within the distance of Virgo cluster \citep{Chilingarian07,Chilingarian10}. However, the only two isolated cEs reported so far -- the one presented here and that of \citet{Huxor13} -- are both located at a greater distance, and so it will be up to future surveys to confirm or refute whether their abundance is really so much lower than that of cEs near massive galaxies.

\section*{Acknowledgments}
This study has made use of NASA's Astrophysics Data System Bibliographic Services and the NASA/IPAC Extragalactic Database (NED). SDSS data were queried from the SDSS archive. Funding for the SDSS/SDSS-III has been provided by the Alfred P. Sloan Foundation, the Participating Institutions, the National Science Foundation, the U.S. Department of Energy, the National Aeronautics and Space Administration, the Japanese Monbukagakusho, the Max Planck Society, and the Higher Education Funding Council for England. The SDSS Web Site is http://www.sdss.org. The SDSS is managed by the Astrophysical Research Consortium for the Participating Institutions. The Participating Institutions are the American Museum of Natural History, Astrophysical Institute Potsdam, University of Basel, University of Cambridge, Case Western Reserve University, University of Chicago, Drexel University, Fermilab, the Institute for Advanced Study, the Japan Participation Group, Johns Hopkins University, the Joint Institute for Nuclear Astrophysics, the Kavli Institute for Particle Astrophysics and Cosmology, the Korean Scientist Group, the Chinese Academy of Sciences (LAMOST), Los Alamos National Laboratory, the Max- Planck Institute for Astronomy (MPIA), the Max-Planck Institute for Astrophysics (MPA), New Mexico State University, Ohio State University, University of Pittsburgh, University of Portsmouth, Princeton University, the United States Naval Observatory, and the University of Washington.

\label{lastpage}

\end{document}